\documentclass[aps,prl,reprint,twocolumn,superscriptaddress,preprintnumbers,nofootinbib,amsmath,amssymb,tighten]{revtex4-1} 

\usepackage{epsfig}
\usepackage{amsfonts}
\usepackage{amsmath}
\usepackage{amssymb}
\usepackage{dsfont}
\usepackage{hyperref}
\usepackage{xspace}
\usepackage{slashed}
\hypersetup{
    colorlinks=true,       
    linkcolor=blue,          
    citecolor=blue,        
    filecolor=blue,      
    urlcolor=blue           
}
\usepackage{xcolor}
\usepackage{color}
\usepackage{graphicx}  %
\usepackage{bm}  %
\usepackage{graphics}

\usepackage{multirow}

\usepackage{caption}
\usepackage{subcaption}
\newcommand{\isovec}[1]{{\vec{#1}}} 
\newcommand{\spacevec}[1]{{\mathbf #1}}

\newcommand{\Nb}{\bar N}

\newcommand{\dslash}[1]{#1 \llap{/\kern-0.5pt}}
\newcommand{\Dslash}[1]{#1 \llap{/\kern+1.5pt}}
\newcommand{\DDslash}[1]{#1 \llap{/\kern+2.3pt}}
\newcommand{\dslashh}[1]{#1 \llap{/\kern+1pt}}

\newcommand{\boldtau}{\mbox{$\isovec \tau$}}

\newcommand{\boldsigma}{\mbox{\boldmath $\sigma$}}

\newcommand{\bea}{\begin{eqnarray}}
\newcommand{\eea}{\end{eqnarray}}
\newcommand{\be}{\begin{equation}}
\newcommand{\ee}{\end{equation}}
\newcommand{\bma}{\begin{pmatrix}}
\newcommand{\ema}{\end{pmatrix}}
\newcommand{\nn}{\nonumber}

\preprint{RIKEN_XXX}

\begin{document}

\title{Strong CP violation in nuclear physics}

\author{Jordy de Vries}
\affiliation{Amherst Center for Fundamental Interactions, Department of Physics, University of Massachusetts Amherst, Amherst, MA 01003, USA}
\affiliation{RIKEN BNL Research Center, Brookhaven National Laboratory, Upton, New York 11973-5000, USA}
\author{Alex Gnech}
\affiliation{University of Pisa, Department of Physics ``E. Fermi", Largo Bruno Pontecorvo, 56127 Pisa, Italy}
\author{Sachin Shain}
\affiliation{Amherst Center for Fundamental Interactions, Department of Physics, University of Massachusetts Amherst, Amherst, MA 01003, USA}

\date{\today}

\begin{abstract}
Electric dipole moments of nuclei, diamagnetic atoms, and certain molecules are induced by $C\!P$-violating nuclear forces. Naive dimensional analysis predicts these forces to be dominated by long-range one-pion-exchange processes, with short-range forces entering only at next-to-next-to-leading order in the chiral expansion. Based on renormalization arguments we argue that a consistent picture of $C\!P$-violating nuclear forces requires a leading-order short-distance operator contributing to ${}^1S_0$-${}^3P_0$ transitions, due to the attractive and singular nature of the strong tensor force in the ${}^3P_0$ channel. The short-distance operator leads to $\mathcal O(1)$ corrections to static and oscillating, relevant for axion searches, electric dipole moments. We discuss strategies how the finite part of the associated low-energy constant can be determined in the case of $C\!P$ violation from the QCD $\bar \theta$ term by the connection to charge-symmetry violation in nuclear systems.

\end{abstract}

\maketitle

\textit{Introduction.} Electric dipole moments (EDMs) of nuclei, atoms, and molecules are excellent probes of new sources of $C\!P$ violation \cite{Chupp:2017rkp}. $C\!P$ violation in the quark and lepton mixing matrices of the standard model (SM) leads to immeasurably small values for EDMs \cite{Seng:2014lea,Yamaguchi:2020eub}, implying that any nonzero measurement is either due to the so-far undiscovered QCD $\bar \theta$ term or from beyond-the-SM (SM) sources of $C\!P$ violation. Current experimental EDM limits \cite{Graner:2016ses,Andreev:2018ayy,Abel:2020gbr} set strong constraints on BSM models with additional $C\!P$-violating phases such as supersymmetry, leptoquarks, multi-Higgs, or left-right symmetric models, and  various scenarios of electroweak baryogenesis \cite{Morrissey:2012db}. In the framework of the SM effective field theory (SMEFT), EDM limits constrain a large set of $C\!P$-odd dimension-six operators at the multi-TeV scale, well above limits from high-energy collider experiments \cite{Cirigliano:2016nyn}.

The interpretation of EDM experiments requires care. It is a non-trivial task to connect EDMs of complex objects such as nuclei or molecules to the underlying $C\!P$-violating source at the quark level. Recent years have seen significant theoretical improvements towards model-independent first-principle calculations of EDMs from a combination of lattice QCD \cite{Bhattacharya:2015esa,Abramczyk:2017oxr,Dragos:2019oxn}, chiral EFT ($\chi$EFT) \cite{Mereghetti:2010tp,deVries:2012ab, Bsaisou:2014oka}, and nuclear calculations \cite{Ban:2010ea,deVries:2011an,Bsaisou:2014zwa,Dobaczewski:2018nim, Gnech:2019dod}. The chain of logic is roughly as follows: the SMEFT framework allows for the derivation of a general set of dimension-four (the QCD $\bar \theta$ term) and -six $C\!P$-violating operators involving light quarks, gluons, and photons. $\chi$EFT, the low-energy EFT of QCD, is used to construct the corresponding $C\!P$-violating interactions among the relevant low-energy degree of freedoms: pions, nucleons, and photons. Each interaction in the chiral Lagrangian comes with a low-energy constant (LEC) that encodes the nonperturbative QCD dynamics that is ideally calculated from lattice QCD (LQCD). EDMs can be then be calculated in terms of the LECs in the CP-odd chiral Lagrangian.

The $\chi$EFT framework provides an expansion of hadronic and nuclear amplitudes in terms of $p/\Lambda_\chi$ where $p \sim k_F  \sim m_\pi \sim \mathcal O(100\,\mathrm{MeV})$ and $\Lambda_\chi \sim 4\pi F_\pi \sim \mathcal O(1\,\mathrm{GeV})$ \cite{Weinberg:1978kz,Hammer:2019poc}, where $F_\pi \simeq 92.4$ MeV is the pion decay constant. The electric dipole form factors of nucleons were calculated up to next-to-next-to-leading order (N${}^2$LO) in the chiral expansion \cite{crewtheretal,Mereghetti:2010kp,Guo:2012vf,Seng:2014pba}. EDMs of nuclei require the derivation of both $C\!P$-conserving and $C\!P$-violating forces and currents. The $C\!P$-odd nucleon-nucleon ($N\!N$) potential was calculated up to N${}^2$LO in Refs.~\cite{Maekawa:2011vs,Gnech:2019dod} and used to calculate EDMs of light nuclei and diamagnetic atoms \cite{Ban:2010ea,deVries:2011an,Bsaisou:2014zwa,Dobaczewski:2018nim, Gnech:2019dod}. 

The derivation of the $C\!P$-odd $N\!N$ potential of Refs.~\cite{Maekawa:2011vs,Gnech:2019dod} is based on Weinberg's power-counting scheme \cite{Weinberg:1990rz}. In this scheme, the $C\!P$-odd potential arises from one-pion-exchange (OPE) diagrams, whose LECs can in principle be fixed from processes involving just nucleons and pions (only in principle as $\pi N$ scattering experiments are not sufficiently accurate). Chiral symmetry does not forbid purely nuclear short-distance interactions with LECs that can only be fixed in nuclear systems. Indeed, in the $C\!P$-conserving potential the leading-order (LO) potential consists of OPE diagrams and two non-derivative contact interactions in ${}^1S_0$ and ${}^3S_1$ waves. In the $C\!P$-violating case, $N\!N$ interactions require at least one space-time derivative and Weinberg's power-counting scheme predicts short-distance operators to enter at N${}^2$LO in the chiral expansion. This is welcome news, as it implies that nuclear EDMs can be calculated in terms of only a few LECs and ratios of EDMs can be used to pinpoint the underlying $C\!P$-violating source \cite{jordyunraveling}.

Weinberg's power counting scheme is based on naive dimensional analysis (NDA) of the $N\!N$ LECs \cite{Manohar:1983md} which is not always reliable for nuclear physics. NDA does not in all cases lead to order-by-order renormalized nuclear amplitudes \cite{Kaplan:1996xu,vanKolck:2020llt}, as required in a consistent EFT. This is most clear in partial waves where OPE is attractive and non-perturbative, such as the ${}^3P_0$ channel, where phase shifts show oscillatory limit-cycle-like cut-off dependence \cite{Nogga:2005hy} that can not be renormalized at LO in Weinberg's scheme. The same problem affects external currents inserted in $N\!N$ scattering states in perturbation theory \cite{Valderrama:2014vra,Cirigliano:2018hja}. 
In this work, we investigate long-distance $C\!P$-violating OPE potentials and demonstrate that renormalization requires a LO short-distance operator for ${}^1S_0$-${}^3P_0$ transitions. This has direct consequences for the interpretation of EDM experiments in terms of the QCD $\bar \theta$ term or higher-dimensional operators, and axion searches via oscillating nuclear EDM experiments \cite{Budker:2013hfa,Stadnik:2013raa}. 

\textit{Setup of the calculation.} We first consider the case of strong $C\!P$ violation from the QCD $\bar \theta$ term. The relevant Lagrangian is given by \cite{'tHooft:1976fv,Baluni:1978rf}
\begin{equation}
\mathcal L =  \bar q i\, \slash \!\!\!\!\mathcal D q - \bar q \left( \mathcal M - i \gamma_5 m_\star \bar \theta \right)q\,,
\end{equation}
where $q=(u\,d)^T$ denotes the quark field, $\mathcal D_\mu$ is the color and electromagnetic covariant derivative, $\mathcal M = \mathrm{diag}(m_u,m_d)$ the quark mass matrix, $m_\star = m_u m_d/(m_u +m_d)$, and the QCD angle $\bar \theta$. The relevant chiral Lagrangian can be constructed with well-known methods \cite{Bernard:1995dp}, and the leading $C\!P$-even and $C\!P$-odd pion-nucleon interactions are given by
\begin{equation}\label{LpiN}
\mathcal L_{\pi N} =  -\frac{g_A}{2F_\pi}\boldsymbol{\nabla}\isovec \pi \cdot \Nb \boldtau\boldsigma N + \bar g_0 \bar N \isovec \pi \cdot \boldtau N + \dots\,,
\end{equation}
in terms of the non-relativistic nucleon doublet $N=(p\,n)^T$ and the pion triplet $\isovec \pi$,  $g_A \simeq 1.27$ is the nucleon axial coupling, and $\bar g_0 =\mathcal O \left(m_\star  \bar \theta/F_\pi\right)$  a $C\!P$-odd LEC. The dots denote interactions involving more pions. The QCD $\bar \theta$ term is related by a chiral rotation to the isospin-breaking component of the quark masses \cite{crewtheretal}, giving a precise determination of $\bar g_0$ \cite{deVries:2015una}
\begin{equation}\label{g0}
\bar g_0 = \frac{\delta m_N^{\mathrm{str}}(1-\varepsilon^2)}{4 F_\pi \varepsilon}\bar \theta = -(14.7 \pm 2.3)\cdot 10^{-3}\,\bar \theta\,,
\end{equation}
where $\delta m_N^{\mathrm{str}}$ is the quark-mass induced part of the proton-neutron mass splitting that has been calculated with LQCD \cite{Brantley:2016our} and $\varepsilon = (m_u-m_d)/(m_u + m_d)$. The value of $\bar g_0$ agrees with a LQCD extraction \cite{Dragos:2019oxn}. 

From the interactions in Eq.~\eqref{LpiN} we calculate the OPE $N\!N$ potentials
\begin{eqnarray}\label{V}
V_{\mathrm{str},\pi} &=& - \frac{1}{(2\pi)^3}\left(\frac{g_A}{2 F_\pi}\right)^2 \boldtau_1\cdot \boldtau_2 \frac{(\boldsigma_1 \cdot \spacevec q)(\boldsigma_2 \cdot \spacevec q)}{\spacevec q^2+m_\pi^2}\,,\nn\\
V_{\bar g_0} &=&-\frac{1}{(2\pi)^3} \frac{g_A \bar g_0}{2 F_\pi}  \boldtau_1\cdot \boldtau_2 \frac{ i (\boldsigma_1 - \boldsigma_2)\cdot \spacevec q}{\spacevec q^2+m_\pi^2}\,,
\end{eqnarray}
where $\spacevec q = \spacevec p - \spacevec p'$ is the momentum transfer between in- and outgoing nucleon pairs with relative momenta $\spacevec p$ and $\spacevec p'$ respectively ($|\spacevec p| = p$ and $|\spacevec p'| = p'$), and $m_\pi$ denotes the pion mass. In addition, we consider $C\!P$-even $N\!N$ interactions in the ${}^1S_0$, ${}^3S_1$, and ${}^3P_0$ waves
\begin{eqnarray}\label{eqC}
V_{\mathrm{str},\mathrm{sd}}&=&  \frac{1}{(2\pi)^3}\left(C_s P_s + C_t P_t + \frac{1}{4}p p'\,C_P P_p\right)\,,
\end{eqnarray}
where $P_{s,t,p}$ project respectively on the ${}^1S_0$, ${}^3S_1$, and ${}^3P_0$ waves. In Weinberg's power counting the $S$-wave contact terms appear at LO while the $P$-wave counter term enters at N${}^2$LO. To obtain the strong $N\!N$ scattering wave functions we solve a Lippmann-Schwinger (LS) equation 
\begin{equation}
T_{\mathrm{str}} = V_{\mathrm{str}} + V_{\mathrm{str}} G_0 T_{\mathrm{str}}\,,\qquad G_0 = (E -  p^2/m_N + i \varepsilon)^{-1}\,,
\end{equation}
with $V_{\mathrm{str}}=(V_{\mathrm{str},\pi} + V_{\mathrm{str},\mathrm{sd}})f_\Lambda(p,p')$, where $f_\Lambda(p,p')$ is a regulator function 
\begin{equation}
f_\Lambda(p,p') = e^{-\left(\spacevec p/\Lambda\right)^4}e^{-\left(\spacevec p'/\Lambda\right)^4}\,,
\label{regulator}
\end{equation}
in terms of a momentum space cut-off $\Lambda$. The LS equation is solved numerically for a wide range of $\Lambda$ to ensure that observables are cut-off independent. 

We briefly discuss results for waves with total angular momentum $j=0,1$ and give explicit results in the Appendix. Solving the LS equation for just the strong OPE potential leads to ${}^1S_0$ and ${}^3S_1$-${}^3D_1$ phase shifts and mixing angles that are cut-off dependent.  In the ${}^3P_1$ and ${}^1P_1$ waves, the strong OPE potential lead to cut-off independent phase shifts that at low energies agree well with experimental data. In the ${}^3P_0$ channel, however, the phase shifts arising from OPE are strongly cut-off dependent and undergo a dramatic limit-cycle like behavior, see Fig.~\ref{app_ct}. In Weinberg's power counting, the regulator dependence of the $S$-wave phase shifts  can be absorbed into the LO counter terms $C_s$ and $C_t$ but there is no counter term for the ${}^3P_0$ channel. Following Ref.~\cite{Nogga:2005hy}, we promote $C_P$ to LO and fit $C_{s,t,p}$ to the phase shifts at a center-of-mass energy $E_{\rm CM}= 5$ MeV. The resulting phase shifts are $\Lambda$ independent for a wide range of energies demonstrating that the strong wave functions are properly renormalized. The LECs $C_{s,t,P}$, of course, show significant $\Lambda$ dependence, but this is of no concern as they are not observable. All results are in agreement with Refs.~\cite{Nogga:2005hy,Song:2016ale}.

Having obtained renormalized scattering states, we insert the $C\!P$-odd potential $V_{\bar g_0}$ which causes ${}^1S_0$-${}^3P_0$ and ${}^3S_1$-${}^1P_1$ transitions. We can treat $V_{\bar g_0}$ to very good accuracy in perturbation theory and write
\begin{equation}\label{Tpert}
T_{\bar g_0} = V_{\bar g_0} + V_{\bar g_0} G_0 T_{\mathrm{str}} +   T_{\mathrm{str}} G_0  V_{\bar g_0}+ T_{\mathrm{str}} G_0 V_{\bar g_0} G_0 T_{\mathrm{str}}\,.
\end{equation}
The on-shell scattering matrix $T = T_{\mathrm{str}} + T_{\bar g_0}$ is related to the $S$ matrix
\begin{equation}
S(E_{\mathrm{CM}}) = 1 - i \pi m_N^{3/2} E^{1/2}_{\rm CM} T(p=p'=\sqrt{E_{\rm CM} m_N})\,,
\end{equation}
where $m_N$ is the nucleon mass.
In the $j=0$ channel we parametrize the S matrix by
\begin{equation}
S_{j=0} =  \begin{pmatrix} e^{2i \delta_{1S_0}} & \epsilon^0_{\rm SP} e^{i \left[\delta_{1S_0} +\delta_{3P_0} \right]} \\
						-\epsilon^0_{\rm SP} e^{i \left[\delta_{1S_0} +\delta_{3P_0} \right]} & e^{2i \delta_{3P_0}} \end{pmatrix}\,,
\end{equation}
where $\epsilon^0_{\rm SP} \sim \bar \theta$ denotes the small ${}^1S_0$-${}^3P_0$ mixing angle. The $j=1$ channel is more complicated due to strong ${}^3S_1$-${}^3D_1$ mixing, and for simplicity we expand in the small $S$-$D$ mixing angle $\epsilon$. Up to $\mathcal O(\epsilon^3)$ corrections we write
\begin{eqnarray}
S_{j=1}  &=&  \begin{pmatrix}  e^{2i \delta_{3S_1}}\cos2\epsilon &  i  e^{i \left[\delta_{3S_1} +\delta_{3D_1}\right]}\sin 2\epsilon  & x_{SP} \\
						 i e^{i \left[\delta_{3S_1} +\delta_{3D_1} \right]}\sin 2\epsilon & e^{2i \delta_{3D_1}}\cos 2\epsilon & x_{DP}  \\
						-x_{SP} & - x_{DP}&   e^{2i \delta_{1P_1}}\end{pmatrix}\,,\nn\\
x_{SP}&=& \left[ \epsilon^1_{\rm SP}	+ i \epsilon\ \epsilon_{\rm DP}  \right] e^{i \left[\delta_{3S_1} +\delta_{1 P_1}\right]} \,,\nn\\	 	
x_{DP}&=& \left[ \epsilon_{\rm DP}	+ i \epsilon\ \epsilon_{\rm SP}^1 \right] e^{i \left[\delta_{3D_1} +\delta_{1 P_1}\right]} \,,
\end{eqnarray}
in terms of two $CP$-odd mixing angles $\epsilon^1_{\rm SP}$ and $\epsilon_{\rm DP}$. $S$ is antisymmetric in the $S$-$P$ and $P$-$D$ elements due to time-reversal violation. The $C\!P$-odd mixing angles $\epsilon^{0,1}_{\rm SP} $ and $\epsilon_{\rm DP} $ are in principle observable in, for example, spin rotation of polarized ultracold neutrons on a polarized hydrogen target \cite{Liu:2006qp}, but it is unlikely that these experiments can reach a sensitivity that is competitive with EDM experiments, although neutron transmission experiments using heavy target nuclei might be up to the task \cite{Fadeev:2019bwc,Schaper:2020akw}. Nuclear EDMs can be written as linear combinations of the mixing angles in addition to contributions from CP-odd electromagnetic currents such as constituent nucleon EDMs.

\begin{figure}[t]
  \begin{subfigure}[b]{1\columnwidth}
    \includegraphics[width=\linewidth]{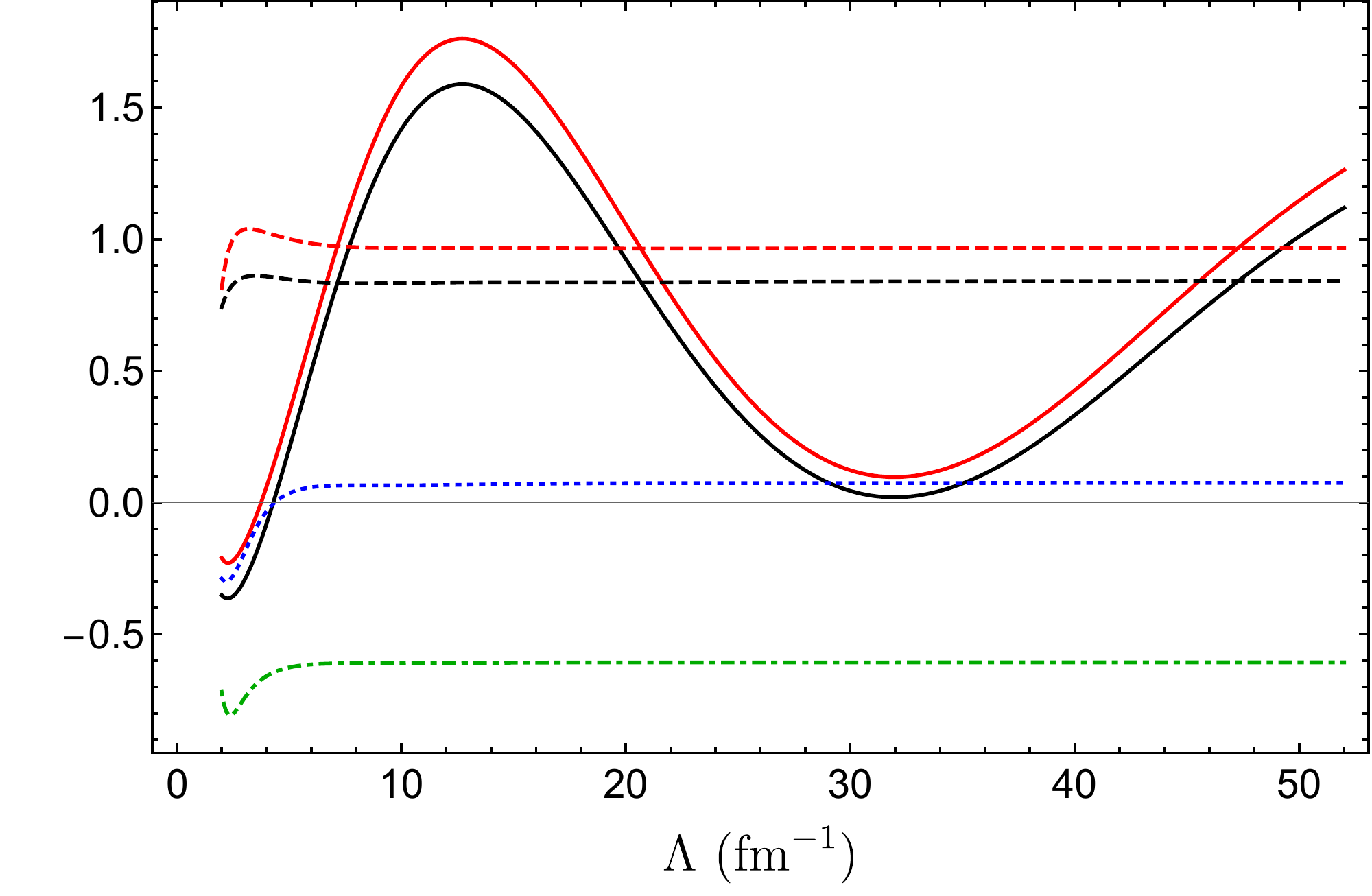}
  \end{subfigure}
  \hfill 
  \\
  \begin{subfigure}[b]{1\columnwidth}
    \includegraphics[width=\linewidth]{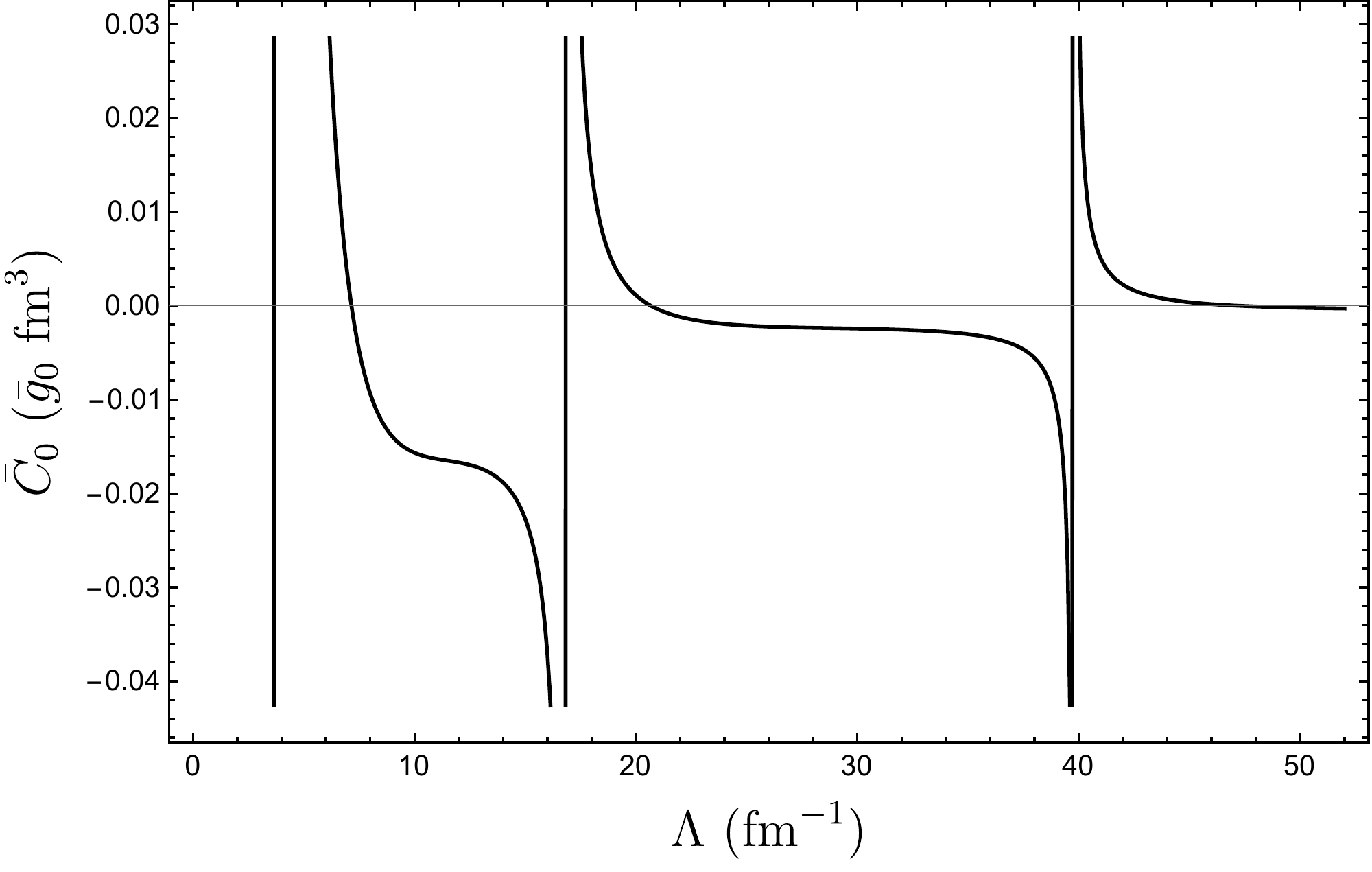}
  \end{subfigure}
  \caption{Top: $\epsilon^0_{\rm SP}$ in units of $\bar g_0$ as a function of $\Lambda$ before (solid) and after (dashed) promoting $\bar C_0$ to LO for $E_{\rm CM}=\{25,\,50\}$ MeV in, respectively, black and red. The blue (dotted) and green (dash-dotted) lines denote, respectively, $\epsilon^1_{\rm SP}$ and $\epsilon_{\rm DP}$ for $E_{\rm CM} = 50$ MeV.  Bottom: $\bar C_0$ as a function of $\Lambda$ . }
  \label{eps0}
\end{figure}

\begin{figure}[t]
    \includegraphics[width=.8\linewidth]{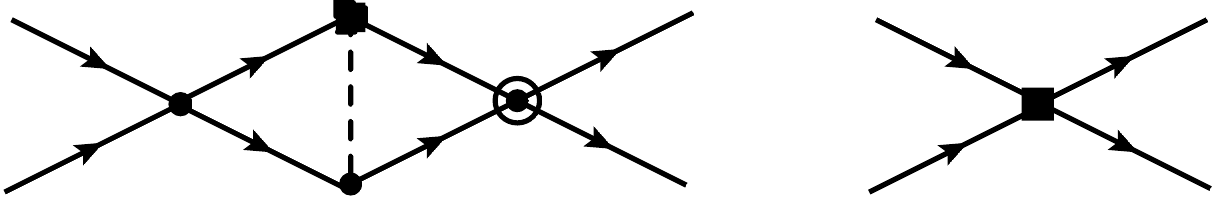}
     \caption{Left: A particular diagram contributing to the regulator dependence of $\epsilon_{\rm SP}^0$. Solid (dashed) lines denote nucleons (pions).  The square denotes an insertion of $\bar g_0$ while the circles denote the $g_A$ or  $C_{s}$ vertices. The circled circle denotes an insertion of $C_P$. Right: short-distance contribution proportional to $\bar C_0$. }
       \label{SPdiagrams}
\end{figure}

The CP-odd mixing angles are observable and should be independent of the value of $\Lambda$. We find that this is the case for $\epsilon_{\rm SP}^{1}$ and $\epsilon_{\rm DP}$ which quickly converge as shown in the top panel of Fig.~\ref{eps0}. However, $\epsilon_{\rm SP}^0$ shows an oscillatory behavior and even changes sign as function of $\Lambda$. There is no sign of convergence whatsoever. We have checked that no regulator dependence appears for any $j=2$ transition after renormalizing the strong $j=2$ scattering states. 
The difference between the behavior of the ${}^1S_0$-${}^3P_0$ and ${}^3\{S,D\}_1$-${}^1P_1$ arises from the absence of a strong counter term in the ${}^1P_1$ channel. The observed regulator dependence arises from divergences in diagrams contributing to $T_{\bar g_0}$ with topology of the left diagram in Fig.~\ref{SPdiagrams}, where $V_{\bar g_0}$ is dressed on both sides by a strong short-distance interaction (an infinite number of related LO diagrams are generated by adding additional strong interactions on either side). At LO this only occurs for ${}^1S_0$-${}^3P_0$ transitions. In $\chi$EFT calculations using Weinberg's power counting, $P$-wave counter terms appear at N${}^2$LO, but are iterated to all orders in the solution of the LS equation \cite{Reinert:2017usi}. Divergent diagrams with the topology of Fig.~\ref{SPdiagrams} reappear and the $C\!P$-odd transitions become regulator dependent. In practice, this might be hard to see numerically as regulators are only varied in a tiny window around $\Lambda =450$ MeV \cite{Bsaisou:2014zwa, Gnech:2019dod}.

\textit{The need for a counter term.} 
The observation that $\epsilon_{\rm SP}^0$ is cut-off dependent implies that $C\!P$-odd nuclear observables that depend on ${}^1S_0$-${}^3P_0$ mixing cannot be directly calculated from $\bar g_0$, and thus $\bar \theta$ via Eq.~\eqref{g0}. An observable that shows regulator dependence in an EFT calculation indicates  there must be an associated counter term that encapsulates missing short-distance physics and absorbs the divergence. In the present context, such counter terms are provided by short-range $C\!P$-odd $N\!N$ interactions, see the right diagram of Fig.~\ref{SPdiagrams}, of the form \cite{deVries:2012ab, Bsaisou:2014oka}
\begin{equation}\label{C12}
\mathcal L_{N\!N} = \bar C_0 \left[ \bar N \boldsigma N\cdot\boldsymbol{\nabla}(\bar N\,N) + \frac{1}{3}\bar N  \vec \tau \boldsigma N\cdot \boldsymbol{\nabla}(\bar N\vec \tau N)\right]\,,
\end{equation}
which projects on ${}^1S_0$-${}^3P_0$. $\bar C_0$ is a LEC that depends on $\Lambda$ in such a way to make $\epsilon_{\rm SP}^0$ $\Lambda$-independent. NDA suggests $\bar C_0 = \mathcal O(m_\star  \bar \theta/(F_\pi^2 \Lambda_\chi^2))$ and a N${}^2$LO contribution, but renormalization enhances $\bar C_0$ by $(4\pi)^2$ making it LO instead.

We now show that promoting  $\bar C_{0}$ to LO indeed renormalizes the ${}^1S_0$-${}^3P_0$ transition. We fit $\bar C_{0}$ at a specific kinematical point to a fictitious measurement of  $\epsilon^0_{SP}$, picking $\epsilon^0_{\rm SP,\rm fit} = 0.01\, \bar g_0$ at $E_{\rm CM} = 5$ MeV for concreteness. The regulator dependence of $\bar C_0$ is shown in the bottom panel of Fig.~\ref{eps0} and shows a limit-cycle-like behaviour driven by $C_P$. The resulting $\epsilon^0_{SP}$ is regulator independent for a wide range of energies as depicted by the dashed lines in the top panel of Fig.~\ref{eps0}. While this method accounts for the regulator-dependent part of the short-distance contributions and renormalizes the $C\!P$-odd amplitude, it cannot account for possible finite contributions from $\bar C_0$. That is, the results in Fig.~\ref{eps0} can shift up or down (they remain flat) if we were to pick different values for $\epsilon^0_{\rm SP,\rm fit}$. The best way to obtain the total short-distance contribution is by fitting to a measurement of $\epsilon^0_{SP}$. This is at present not possible, and even if there was data it would not be satisfactory. We would like to use such data to extract a value of $\bar \theta$.

\textit{Fixing the value of the short-distance LEC.} 
We discuss two potential methods to obtain a value for $\bar C_0$ in the absence of data. The first one is to perform a LQCD calculation of $N\!N \rightarrow N\!N$ scattering in the presence of a nonzero $\bar \theta$ background. There have been significant recent developments in calculations of the nucleon EDM arising from the QCD $\bar \theta$ term by applications of the gradient flow \cite{Luscher:2010iy,Dragos:2019oxn}, and the same techniques could be used to study four-point functions in a $\bar \theta$ vacuum. A major challenge will be to control the signal-to-noise. Already for $C\!P$-conserving $N\!N \rightarrow N\!N$ scattering, signal-to-noise considerations demand pion masses well above the physical point \cite{Orginos:2015aya}. Going to smaller pion masses is even more daunting in case of $C\!P$ violation from the $\bar \theta$ term, as the signal scales as $\sim \bar \theta m_\pi^2$. If such a LQCD calculation is possible, we can obtain $\bar C_0$ from a matching calculation of $\chi$EFT to the lattice data after taking the appropriate continuum and infinite-volume limits. 

On a shorter time-scale a more promising approach is to apply chiral-symmetry relations between the $\bar \theta$ term and the quark masses similar to the relation between $\bar g_0$ and $\delta m^{\rm str}_N$ in Eq.~\eqref{g0}. Using $SU(2)_L \times SU(2)_R$ $\chi$EFT, the operators in Eq.~\eqref{C12} arise from the structures
\begin{eqnarray}
\mathcal L_{N\!N} &=&-\frac{i C_0}{8} \mathrm{Tr} [\chi_-]\bigg[\bar N \boldsigma N\cdot \boldsymbol{\nabla}(\bar NN) \nn\\
 &&\qquad \qquad  +  \frac{1}{3}\bar N  \vec \tau \boldsigma N\cdot \boldsymbol{\nabla}(\bar N\vec \tau N)\bigg],
\end{eqnarray}
where $\chi_- = u^\dagger \chi u^\dagger - u \chi^\dagger u$, $u = \exp(i \vec \tau \cdot \vec \pi/(2 F_{\pi}))$, $\chi = 2 B(\mathcal M + i m_\star \bar \theta)$, and $B= -\langle\bar q q\rangle/F_\pi^2$ related to the chiral condensate. Expanding out the trace gives $\bar C_{0} = (B m_\star \bar \theta) C_{0}$ and a relation to the $C\!P$-conserving but isospin-breaking $N\!N\pi$  operators \cite{Maekawa:2011vs}
\begin{eqnarray}\label{NNpi}
\mathcal L_{N\!N,\pi} &=& \frac{C_0 B (m_d-m_u)}{2}\frac{\pi_0}{F_\pi}\bigg[\bar N \boldsigma N\cdot \boldsymbol{\nabla}(\bar N\,N) \nn\\
&&\qquad \qquad  +  \frac{1}{3}\bar N  \vec \tau \boldsigma N\cdot \boldsymbol{\nabla}(\bar N\vec \tau N)\bigg]\,.
\end{eqnarray}
These operators contribute to charge-symmetry-breaking (CSB) in $N\!N \rightarrow N\!N\pi$ processes \cite{vanKolck:2000ip,Gardestig:2004hs,Nogga:2006cp,Baru:2013zpa}. One of the LO contributions to this CSB process arises from the $N\pi\pi$ vertex related to $\delta m_N^{\rm str}$ by chiral symmetry
\begin{equation}\label{CSB}
\mathcal L_{\rm CSB} = - \frac{\delta m_N^{\mathrm{str}}}{4 F_\pi^2} \bar N \vec \tau \cdot \vec \pi\,\pi_0 N\,.
\end{equation}
 The contact operator in Eq.~\eqref{NNpi} contributes at N${}^2$LO in Weinberg's counting (in agreement with Ref.~\cite{Nogga:2006cp} that relegate counter terms to N${}^4$LO in an expansion in $\sqrt{p/\Lambda_\chi}$).  At the pion threshold, where final-state $\pi N$ interactions can be neglected, the transition operator for the process ${}^1S_0$-${}^3P_0 + \pi$ due to Eq.~\eqref{CSB} is of the same form as $V_{\bar g_0}$. As such, the regulator dependence seen in Fig.~\ref{eps0} appears and $C_0$ must be promoted to LO for renormalization. Unfortunately the simplest process where CSB data is available, $p n \rightarrow d \pi^0$, is not sensitive to $C_0$ due to the isosinglet nature of the deuteron. This motivates 
an investigation of $d d \rightarrow \alpha \pi^0$ using renormalized $\chi$EFT to fit $C_0$  to CSB data \cite{Adlarson:2014yla}, which would provide a determination of $\bar C_{0} = (B m_\star \bar \theta) C_{0}$.

\textit{Consequences for other sources of $C\!P$ or $P$ violation.} 
At the dimension-six level in the SMEFT there appear other $C\!P$-odd sources involving light quarks. The most relevant operators for the present discussion are quark chromo-EDMs and chiral-breaking four-quark operators, which are induced in a wide range of BSM models \cite{Pospelov:2005pr,jordyunraveling}. 
In addition to the isoscalar $\bar g_0$ term in Eq.~\eqref{LpiN}, the LO $C\!P$-odd chiral Lagrangian contains an isovector term
\begin{equation}\label{LpiN2}
\mathcal L_{\pi N} =  \bar g_1 \bar N\pi_0  N \,.
\end{equation}
A potential isotensor term is subleading for all dimension-six operators \cite{deVries:2012ab}. In combination with the strong $g_A$ vertex, an OPE involving $\bar g_1$ causes ${}^1S_0$-${}^3P_0$ and ${}^3S_1$-${}^3P_1$ transitions. Strong ${}^3P_1$ interactions arise solely from long-distance OPE such that the divergent diagrams in Fig.~\ref{SPdiagrams} do not appear and we expect no regulator dependence for ${}^3S_1$-${}^3P_1$ transitions. This is confirmed by explicit calculations. The $j=0$ transition, up to an isospin factor, shows the same regulator dependence as the $\bar g_0$ case and thus a LO isospin-breaking counter term is needed. The associated operator takes the form
\begin{equation}\label{C12}
\mathcal L_{N\!N} = \bar C_1\left[\bar N \tau^3 \boldsigma N\cdot\boldsymbol{\nabla}(\bar N\,N) + \bar N  \boldsigma N\cdot \boldsymbol{\nabla}(\bar N \tau^3  N)\right]\,,
\end{equation}
which projects unto ${}^1S_0$-${}^3P_0$, but only for the neutron-neutron and proton-proton case. The simplest EDM that depends on $\bar g_1$ is the deuteron EDM \cite{Khriplovich:1999qr}, which is targeted in storage-ring experiments \cite{Abusaif:2019gry}. Due to the isosinglet nature of the deuteron, its EDM only depends on ${}^3S_1$-${}^3P_1$ transitions which do not require a counter term for renormalization. There is no such selection rule for more complex EDMs such as ${}^3$He, ${}^{199}$Hg, or ${}^{225}$Ra \cite{Dzuba:2009kn,deVries:2011an,Bsaisou:2014zwa, Dobaczewski:2018nim,Gnech:2019dod,Yanase:2020agg}, and $\bar C_{1}$ must be included at LO. 

Finally, the finiteness of  ${}^3S_1$-${}^3P_1$ transitions is relevant for the field of hadronic parity ($P)$ violation \cite{Haxton:2013aca}. The LO $P$-odd, but $C\!P$-even, chiral Lagrangian induced by $P$-odd four-quark operators contains a single $\pi$N term \cite{Kaplan:1992vj}, usually parametrized as $ (h_\pi/\sqrt{2})\bar N (\vec \pi \times \vec \tau)^3  N$ that in combination with $g_A$ leads to ${}^3S_1$-${}^3P_1$ transitions \cite{Zhu:2004vw,deVries:2020iea}. We have checked explicitly that no regulator dependence appears and no counter terms are needed. The value of $h_\pi$ that has been recently determined from $P$-violating asymmetries in $\vec n p \rightarrow d \gamma$ \cite{Blyth:2018aon}, can thus be directly applied in calculations of other $P$-odd observables.

\textit{Conclusion.} We have argued the need for a leading-order short-range $C\! P$-violating counter term in ${}^1S_0$-${}^3P_0$ transitions that affects calculations of EDMs and $C\!P$ violation in neutron-nucleus scattering at the $\mathcal O(1)$ level. This directly affects the interpretation of experimental limits, and hopefully future signals, in terms of the QCD $\bar \theta$ term and other $C\!P$-odd sources, and the interpretation of axion dark matter searches via oscillating EDMs. For $C\!P$ violation from the $\bar \theta$ term, we have proposed strategies to obtain the value of the associated low-energy constant, $\bar C_0$, from existing data on charge-symmetry-breaking in pion production in few-body systems. We hope our results stimulate determinations of $\bar C_1$ using lattice QCD and analyses of CSB data, and calculations of the impact of the short-range operator on observables of experimental interest such as (oscillating) EDMs and time-reversal-odd scattering observables.

\begin{acknowledgments}
We would like to thank Emanuele Mereghetti and Bira van Kolck for valuable discussions. We thank Nodoka Yamanaka for discussions in the initial stage of this work. JdV is supported by the 
RHIC Physics Fellow Program of the RIKEN BNL Research Center.
\end{acknowledgments}

\bibliography{references}

\appendix

\section{Renormalization of the strong scattering states}

We consider $N\!N$ scattering in the center-of-mass frame with energy $E_{CM}$. The momenta  of the incoming (outgoing) nucleons and their quantum numbers are denoted by $p$  ($p'$) and $\alpha=\alpha((ls)jm_j,tm_t)\ (\alpha')$, respectively. $l$, $s$, $j$, $m_j$, $t$, $m_t$ denote, respectively, orbital angular momentum, spin, total angular momentum, third component of total angular momentum, total isospin, and third component of isospin. We focus on the LO $C\!P$-even potential $V_{\mathrm{str}}$ as given by Weinberg's power counting
\begin{align}
V_{\mathrm{str},\pi} &= - \frac{1}{(2\pi)^3}\left(\frac{g_A}{2 F_\pi}\right)^2 \boldtau_1\cdot \boldtau_2 \frac{(\boldsigma_1 \cdot \spacevec q)(\boldsigma_2 \cdot \spacevec q)}{\spacevec q^2+m_\pi^2}\,,
\\
V_{\mathrm{str},\mathrm{sd}}&=  \frac{1}{(2\pi)^3}\left(C_s P_s + C_t P_t\right).
\end{align}
The LS equation is given by
\begin{align}
T^{\alpha'\alpha}_{\mathrm{str}}(p',p,E_{CM}) &=
V_{\mathrm{str}}^{\alpha'\alpha}(p',p)
\nn
\\
+
\sum_{\alpha''}
\int dp''\,p''^2
V_{\mathrm{str}}^{\alpha'\alpha}&(p',p'')
G_0(p''^2)
T^{\alpha''\alpha}_{\mathrm{str}}(p'',p,E_{CM})\,,
\nn
\end{align}
which we solve numerically after introducing the regulator function  $f_\Lambda(p,p')$ in Eq.~\eqref{regulator}.

\begin{figure}[t]
  \begin{subfigure}[b]{.85\columnwidth}
    \includegraphics[width=.85\linewidth]{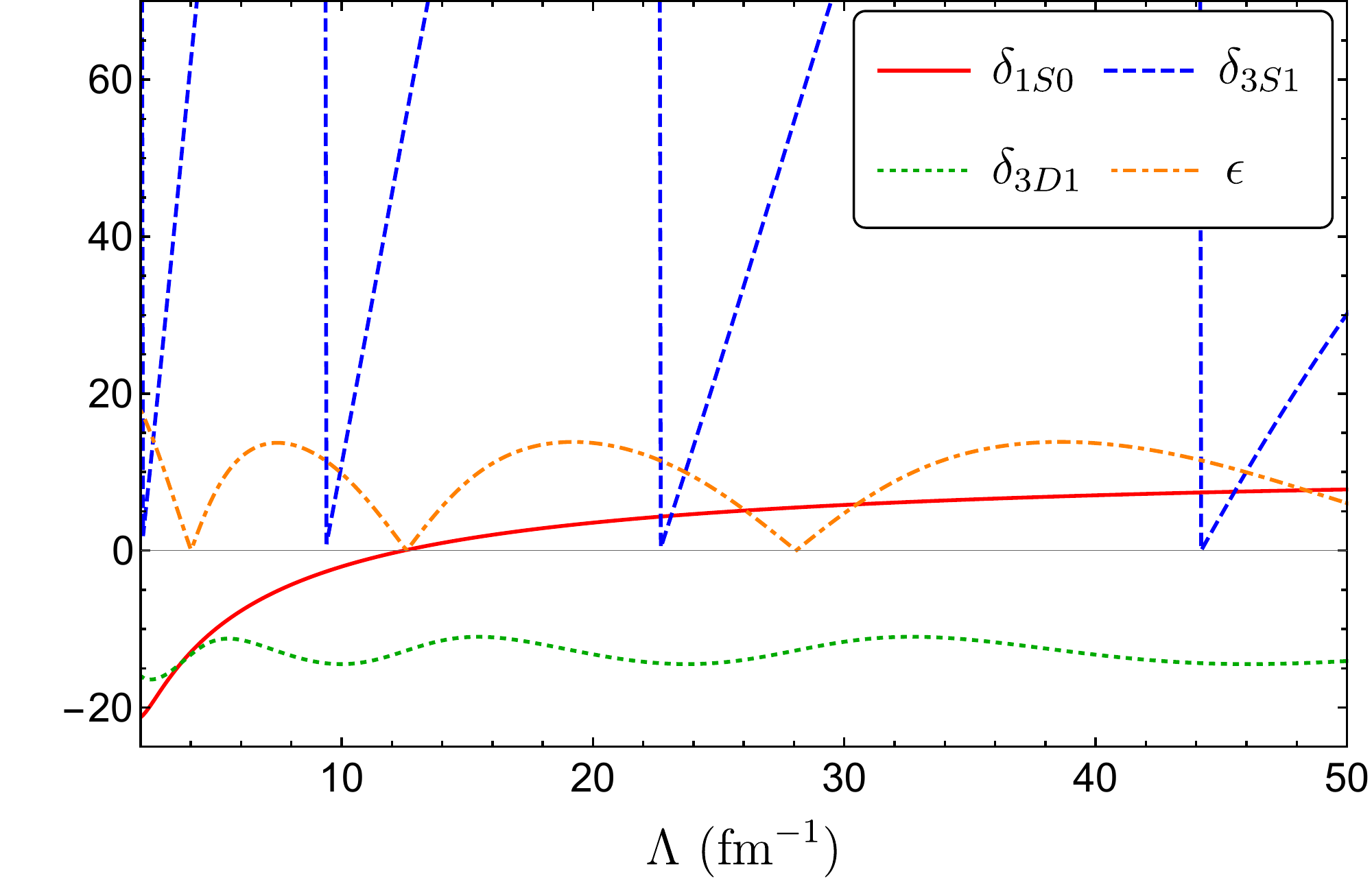}
  \end{subfigure}
  \hfill 
  \\
  \begin{subfigure}[b]{0.85\columnwidth}
    \includegraphics[width=.85\linewidth]{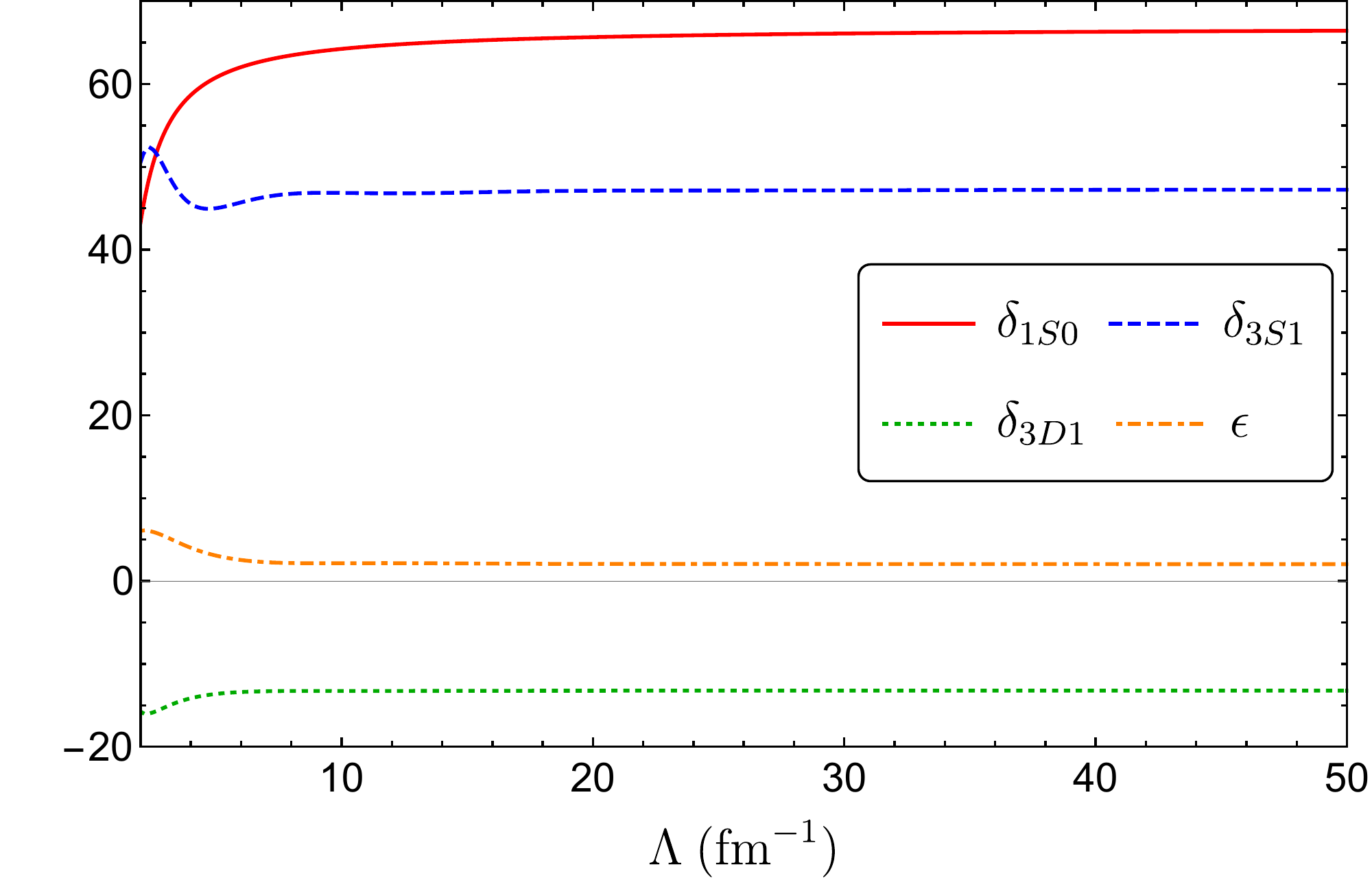}
  \end{subfigure}
  \caption{Phase shifts and mixing angle for the ${}^1S_0$ and ${}^3S_1$-${}^3D_1$ channels at $E_{\rm CM} = 50$ MeV from just the strong OPE potential (top) and after renormalization (bottom) as a function of the  regulator $\Lambda$.}
  \label{app_SD}
\end{figure}

\begin{figure}[t]
  \begin{subfigure}[b]{.85\columnwidth}
    \includegraphics[width=.85\linewidth]{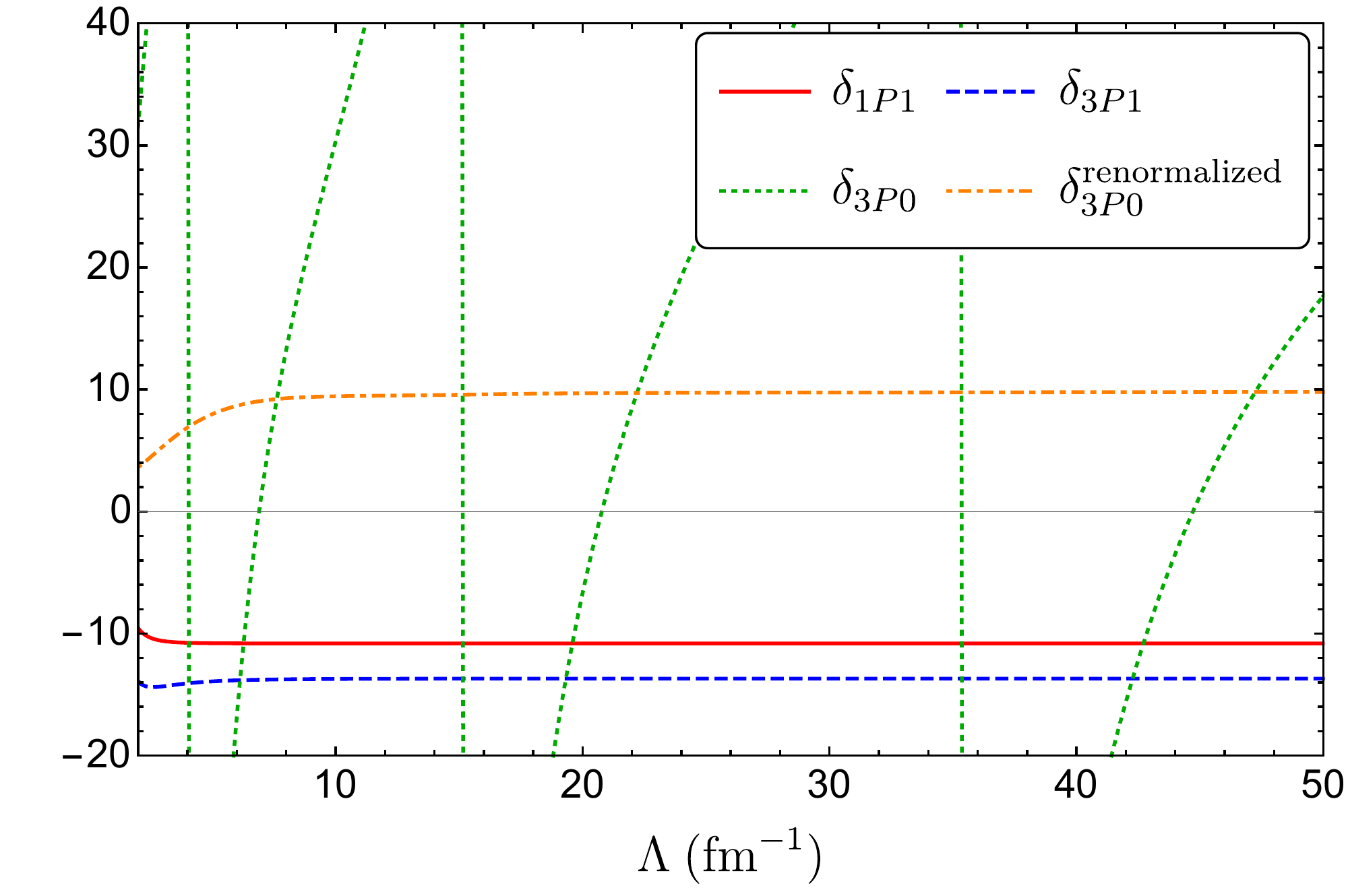}
  \end{subfigure}
  \hfill 
  \\
  \begin{subfigure}[b]{0.85\columnwidth}
    \includegraphics[width=.85\linewidth]{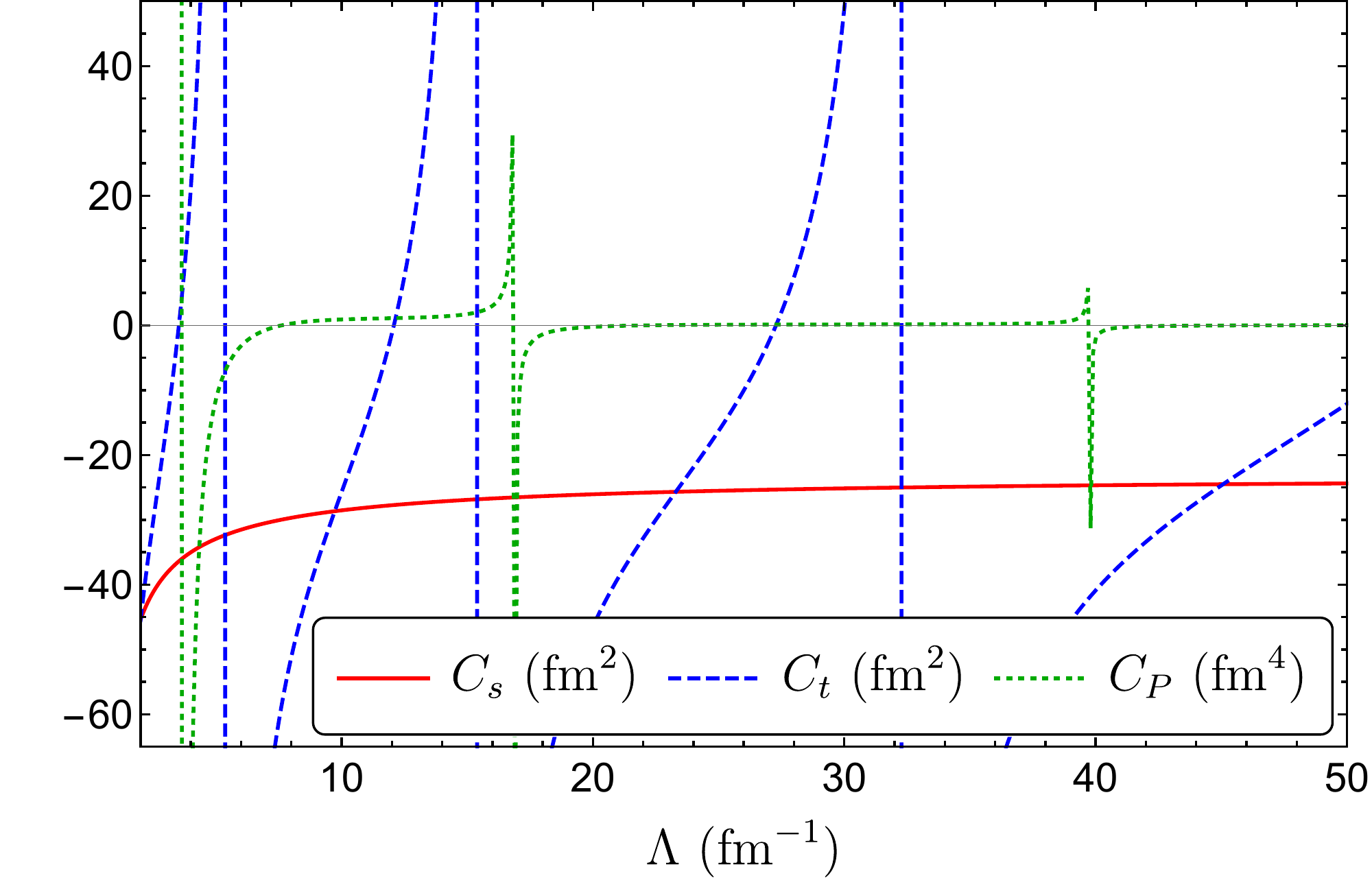}
  \end{subfigure}
  \caption{top: ${}^3P_0$, ${}^3P_1$, and ${}^1P_1$ phase shifts at $E_{\rm CM} = 50$ MeV as a function of the  regulator $\Lambda$. The green (dotted) line denotes the ${}^3P_0$ phase shifts in Weinberg's power counting where no counter term is available for renormalization. The orange (dash-dotted) is the result after promoting $C_P$ to LO.
   Bottom: Low-energy constants $C_s$, $C_t$, and $C_P$ as a function of $\Lambda$.}
  \label{app_ct}
\end{figure}

The phase shifts and mixing angles calculated using just the OPE potential are cut-off dependent in the ${}^1S_0$ and ${}^3S_1$-${}^3D_1$ channels, see the top panel of  Fig. \ref{app_SD}. This is resolved by including the short-distance counter terms $C_s$ and $C_t$ acting in the $^1S_0$ and $^3S_1$ waves. We fit the LECs to reproduce the strong phases shifts at $E_{CM}= 5 $ MeV. The phase shifts then become cut-off independent for a wide range of energies, as exemplified for $E_{CM}= 50$ MeV in the bottom panel of Fig. \ref{app_SD}. The regulator dependence of $C_s$ and $C_t$ is given in the bottom panel of Fig.~\ref{app_ct}. 

Using just the strong OPE potential leads to cut-off independent phase shifts in the ${}^1P_1$ and ${}^3P_1$ channels, see the top panel of Fig.~\ref{app_ct}. In the ${}^3P_0$ wave, however, the strong tensor force is attractive leading to phase shifts that are very sensitive to short-distance physics and the phase shifts show a limit-cycle behaviour as a function of $\Lambda$. Unlike for the ${}^1S_0$ and ${}^3S_1$ channels there does not appears a counter term that can absorb this regulator dependence in Weinberg's power counting.  We therefore promote the ${}^3P_0$ counter term with LEC $C_P$ in Eq.~\eqref{eqC} 
to LO and fit $C_P$ to the $^3P_0$ phase-shift at $E_{CM}=5$ MeV. With this modified power counting the phase-shifts becomes cut-off independent, see top panel of Fig. \ref{app_ct}. The regulator dependence of $C_P$ is given in the bottom panel of Fig.~\ref{app_ct}. 

\end{document}